\documentclass[pra,onecolumn,floatfix,superscriptaddress,longbibliography,notitlepage, nofootinbib]{revtex4-1}
\pdfoutput=1

\usepackage[utf8]{inputenc}
\usepackage{braket}
\usepackage{amsmath}
\DeclareMathOperator{\Tr}{Tr}
\usepackage{dcolumn}   
\usepackage{bm}        
\usepackage{amssymb}
\usepackage{mathtools}
\usepackage{tikz}
\usepackage{qcircuit}
\usepackage{appendix}
\usepackage{hyperref}
\usepackage{subcaption}
\usepackage{amsmath,amsfonts,amssymb}
\usepackage{mathtools}
\usepackage{thmtools,thm-restate}
\usepackage{algorithm}
\usepackage{mathrsfs}
\usepackage{tikz}
\usepackage{subcaption}
\usepackage{pgfplots}
\usetikzlibrary{3d}
\usepackage{tkz-euclide}

\usepackage{algpseudocode}

\usepackage{ulem}

\newcommand{\bbf}{\textbf{b}}
\newcommand{\xbf}{\textbf{x}}

\newcommand{\Ibb}{\mathbb{I}}

\newcommand{\Rbb}{\mathbb{R}}

\newtheorem{definition}{Definition}
\newtheorem{theorem}{Theorem}
\newtheorem{lemma}{Lemma}

\usetikzlibrary{arrows.meta}

\def\BibTeX{{\rm B\kern-.05em{\sc i\kern-.025em b}\kern-.08em
    T\kern-.1667em\lower.7ex\hbox{E}\kern-.125emX}}

\begin{document}
\title{New Quantum Algorithm For Solving Linear System of Equations}
\author{Nhat A. Nghiem}
\email{nhatanh.nghiemvu@stonybrook.edu}
\affiliation{Department of Physics and Astronomy, State University of New York at Stony Brook, Stony Brook, NY 11794-3800, USA}
\affiliation{C. N. Yang Institute for Theoretical Physics, State University of New York at Stony Brook, Stony Brook, NY 11794-3840, USA}

\begin{abstract}
Linear equations play a pivotal role in many areas of science and engineering, making efficient solutions to linear systems highly desirable. The development of quantum algorithms for solving linear systems has been a significant breakthrough, with such algorithms rapidly becoming some of the most influential in quantum computing. Subsequent advances have focused on improving the efficiency of quantum linear solvers and extending their core techniques to address various challenges, such as least-squares data fitting. In this article, we introduce a new quantum algorithm for solving linear systems based on the gradient descent method. Inspired by the vector/density state formalism, we represent a point, or vector, as a density state-like entity, enabling us to leverage recent advancements in quantum algorithmic frameworks, such as block encoding, to directly manipulate operators and carry out the gradient descent method in a quantum setting. The operator corresponding to the intermediate solution is updated iteratively, with a provable guarantee of convergence. The quantum state representing the final solution to the linear system can then be extracted by further measurement. We provide a detailed complexity analysis and compare our approach with existing methods, demonstrating significant improvement in certain aspects. 
\end{abstract}
\maketitle

\section{Introduction}
Quantum computing holds great promise in solving problems that are beyond the reach of classical computers. Indeed, numerous examples across various domains have demonstrated quantum advantages. Grover's algorithm \cite{grover1996fast} achieves a quadratic speed-up in searching an unstructured database. Shor's algorithm \cite{shor1999polynomial}, along with recent improvements \cite{regev2023efficient}, exhibits a superpolynomial speed-up in integer factorization. Deutsch's algorithm \cite{deutsch1985quantum,deutsch1992rapid} introduced a black-box computation model, demonstrating that a quantum computer can determine a black-box's properties with minimal queries. In a similar vein, Jordan \cite{jordan2005fast} showed that a single evaluation of a black-box function suffices to estimate its gradient. Other works \cite{aharonov2006polynomial, lloyd2016quantum, ubaru2021quantum, nghiem2023constant, nghiem2023quantum} have explored various quantum speed-ups for computational problems in algebraic topology. In a more physically motivated setting, a series of studies have demonstrated quantum computers' ability to efficiently simulate a variety of quantum systems \cite{lloyd1996universal, berry2007efficient, berry2012black, berry2014high, berry2015hamiltonian, childs2010relationship, childs2018toward, gerritsma2010quantum, berry2020time, chen2021quantum, an2022time, low2017optimal, low2019hamiltonian}.

Among the many quantum algorithms that have been proposed—and many more that will certainly emerge—the Harrow-Hassidim-Lloyd (HHL) algorithm \cite{harrow2009quantum} for solving linear systems stands out as one of the most influential. Beyond merely providing a quantum state corresponding to the solution of a linear system, the authors of \cite{harrow2009quantum} demonstrated that solving linear systems is $\textbf{BQP}$-complete. This implies that, under widely accepted complexity-theoretic assumptions, classical computers cannot achieve a fundamentally faster solution. As shown in \cite{harrow2009quantum}, the ability to solve linear systems in logarithmic time (with respect to problem dimension) opens up exciting practical applications for quantum computing, as linear systems are ubiquitous in science and engineering. The techniques introduced in the HHL algorithm have also inspired other quantum algorithms, such as quantum support vector machines \cite{rebentrost2014quantum} and quantum data fitting algorithms \cite{wiebe2012quantum}.

Building on the success of the HHL algorithm, several attempts \cite{clader2013preconditioned, childs2017quantum, kerenidis2020quantum, huang2019near, bravo2023variational, wossnig2018quantum, subacsi2019quantum} have sought to improve it in various ways. The HHL algorithm relies on key subroutines, including Hamiltonian simulation techniques \cite{berry2007efficient, berry2012black} and quantum phase estimation \cite{kitaev1995quantum, abrams1999quantum}. However, as noted in \cite{childs2017quantum}, quantum phase estimation leads to a linear dependence on the inverse of the error tolerance, making further optimizations necessary. To address this, the authors of \cite{childs2017quantum} employed alternative approximation techniques—namely, Fourier and Chebyshev approximations—resulting in a quantum linear solver with polylogarithmic scaling in reciprocal error. The work in \cite{clader2013preconditioned} tackled the issue of high condition numbers in real-world linear systems by incorporating a preconditioning technique, which reformulates the system into an equivalent one with a provably bounded condition number while maintaining the same solution. Ref.\cite{subacsi2019quantum} introduced a quantum linear solver based on adiabatic quantum computing, achieving the same complexity as HHL. Meanwhile, Refs.\cite{wossnig2018quantum, kerenidis2020quantum} developed quantum linear solvers leveraging the augmented quantum random access memory model, targeting dense linear systems and achieving a quadratic speed-up over both HHL and \cite{childs2017quantum} in the same setting. Additionally, \cite{bravo2023variational, huang2019near} proposed quantum linear solvers tailored for near-term quantum devices, bypassing the high-depth requirements of other approaches such as \cite{harrow2009quantum, childs2017quantum}.

In this work, we introduce a new quantum algorithm for solving linear systems. Our input model is similar to that of the HHL algorithm, in which oracle access to the matrix entries is provided. Our approach reformulates the problem as an optimization task, which is then solved using gradient descent—a strategy explored in \cite{huang2019near, kerenidis2020quantum}. However, \cite{huang2019near} employs a variational method, which is more heuristic, while \cite{kerenidis2020quantum} assumes a different input model based on augmented quantum random access memory. We show that, by defining an appropriate cost function, its gradient can be computed analytically. Furthermore, instead of representing a point in Hilbert space with a state vector (e.g., $\mathbf{x}$), we represent it using a density matrix-like operator $\mathbf{x} \mathbf{x}^\dagger$. This alternative representation enables us to leverage the recently introduced quantum singular value transformation (QSVT) framework \cite{gilyen2019quantum} to perform gradient descent on $\mathbf{x} \mathbf{x}^\dagger$. Although this representation deviates from the conventional notion of a state vector, we demonstrate that the quantum state corresponding to the  solution of the optimization problem, and thus of the linear system, can still be recovered from the density operator representation. Thus, beyond providing a novel quantum approach for solving linear systems, our work also highlights the versatility of QSVT, adding to its growing list of applications \cite{gilyen2019distributional, chakraborty2018power}. We note that while \cite{gilyen2019quantum} also proposed using QSVT to solve linear systems, their approach relies directly on polynomial approximations, making it technically distinct from ours.

Our work is organized as follows. In the next Section \ref{sec: overview}, we provide an overview, including the main objective, key input assumptions, summary of prior results and preliminaries for our method. Section \ref{sec: mainframework} is devoted to outline our main proposal, which is a new quantum algorithm for solving linear system. A detailed analysis of algorithm, including error and time complexity is provided Section \ref{sec: analysis}. Conclusion and outlook is left for Section \ref{sec: conclusion}.

\section{Overview and Some Preliminaries}
\label{sec: overview}
\noindent
\textbf{Main Objective and Assumptions.} A linear system of size $n \times n$ is defined as follows.
\begin{align}
    A\xbf = \bbf
\end{align}
where $A$ is a matrix of size $n \times n$ and $\bbf$ is a $n$-dimensional vector. If $A$ is non-singular, i.e., having full rank, then the solution is given by $\xbf = A^{-1} \bbf$, and it is unique. For $A$ being singular, then the inverse of $A$ doesn't exist, and in principle there are many solutions. Similar to \cite{harrow2009quantum}, we assume in our work that $A$ is non-singular. Additionally, we assume that $A$ is Hermitian and is $s$ sparse, i.e., for each row/column of $A$, there are at most $s$ non-zero entries. The operator norm of $A$ is assumed to be less than 1, or equivalently, the eigenvalues of $A$ are bounded between $(1/\kappa,1)$, for which $\kappa$ is called the conditional number of $A$. We remark that the aforementioned conditions are without loss of generality. If $A$ is not Hermitian, as also mentioned in \cite{harrow2009quantum}, there is a Hermitian embedding trick:
\begin{align}
    A \longrightarrow  A' = \begin{pmatrix}
    \textbf{0}_{n\times n} & A\\
    A^\dagger & \textbf{0}_{n \times n}
\end{pmatrix}
\end{align}
and instead of solving $A\xbf = \bbf$, we can solve:
\begin{align}
    A' \begin{pmatrix}
    \textbf{0}_n\\
    \xbf 
\end{pmatrix} = \begin{pmatrix}
    \bbf \\
    \textbf{0}_n
\end{pmatrix}
\end{align}
where $\textbf{0}_n, \textbf{0}_{n \times n}$ is an $n$-dimensional vector, matrix having 0 as entries, respectively. The matrix $A'$ is Hermitian, and solving the above system can produce $\xbf$. Regarding the operator norm of $A$, we can always rescale the matrix $A$ so that its eigenvalues are less than $1$. The final assumptions we make are that we are provided with an oracle/black-box access to the entries of $A$, and that $\bbf$ is a unit vector with a known preparation quantum circuit.  

In a quantum format, the goal is not to specifically solve for $\xbf$, but rather to obtain a quantum state $\ket{\Tilde{\xbf}}$ that is close to $\ket{\xbf} = A^{-1} \bbf/ ||A^{-1} \bbf||$, i.e., $ |\ket{\Tilde{\xbf}} - \ket{\xbf}| \leq \delta$. As we mentioned earlier, there are quite a few quantum linear solving algorithms; however, only \cite{harrow2009quantum, childs2017quantum} and \cite{clader2013preconditioned} fall into the same input model as ours. For a glimpse of our contribution, we provide the following table summarizing all results.
\begin{table}[H]
    \centering
    \begin{tabular}{|c|c|c|c|}
    \hline
    Our method  & Ref.~\cite{harrow2009quantum} & Ref.~\cite{childs2017quantum} & Ref.~\cite{clader2013preconditioned}  \\
    \hline
    $\mathcal{O}\Big( s^2 \frac{1}{\delta}  \big( \log(n) + s^2 \big) \log^{3.5} \frac{s}{\delta} \Big)$  
    & $ \mathcal{O}\Big(\frac{1}{\delta} s \kappa \log n \Big) $ 
    & $\mathcal{O}\Big( s \kappa^2  (\log(n) + \operatorname{poly} \log \frac{\kappa}{\delta} )\Big)$ 
    & $\mathcal{O}\Big( \frac{1}{\delta} \cdot  s^7 \Big)$  
    \\
    \hline
    \end{tabular}
    \caption{Table summarizing our result and relevant works. }
    \label{tab:my_label}
\end{table}
The above table indicates that our algorithm is well-suited for linear system with high conditional numbers. Comparing to \cite{clader2013preconditioned}, there is a power-of-five improvement in the sparsity, which is a major improvement.

\noindent
\textbf{Preliminaries.} In the following, we summarize some important definition and main recipes of our subsequent construction. We keep their statements brief but precise for simplicity, with their proofs/ constructions referred to in their original works \cite{gilyen2019quantum}. 

\begin{definition}[Block Encoding Unitary]~\cite{low2017optimal, low2019hamiltonian, gilyen2019quantum}
\label{def: blockencode} 
Let $A$ be some Hermitian matrix of size $N \times N$ whose matrix norm $|A| < 1$. Let a unitary $U$ have the following form:
\begin{align*}
    U = \begin{pmatrix}
       A & \cdot \\
       \cdot & \cdot \\
    \end{pmatrix}.
\end{align*}
where $\cdot$ refers to other possibly non-zero entry matrix blocks. We use this bullet notation to avoid confusion with $(\cdot)$ notation used in the main text that refers to $0$ entry. Then $U$ is said to be an exact block encoding of matrix $A$. Equivalently, we can write:
\begin{align*}
    U = \ket{ \bf{0}}\bra{ \bf{0}} \otimes A + \cdots,
\end{align*}
or alternatively, $A = \bra{\bf 0}\otimes \Ibb \ U \ \ket{\bf 0}\otimes \Ibb$
where $\ket{\bf 0}$ refers to the ancilla system required for the block encoding purpose. In the case where the $U$ has the form 
$$ U  =  \ket{ \bf{0}}\bra{ \bf{0}} \otimes \Tilde{A} + \cdots, $$
where $|| \Tilde{A} - A || \leq \epsilon$ (with $||.||$ being the matrix norm), then $U$ is said to be an $\epsilon$-approximated block encoding of $A$.
\end{definition}

The above definition has multiple natural corollaries. First, an arbitrary unitary $U$ block encodes itself. Suppose $A$ is block encoded by some matrix $U$, then $A$ can be block encoded in a larger matrix by simply adding ancillas (which have dimension $m$). Note that $\Ibb_m \otimes U$ contains $A$ in the top-left corner, which is a block encoding of $A$ again by definition. Further, it is almost trivial to block encode the identity matrix of any dimension. For instance, we consider $\sigma_z \otimes \Ibb_m$ (for any $m$), which contains $\Ibb_m$ in the top-left corner. 

From the above definition, suppose $\ket{\phi}$ is arbitrary state having the same dimension as $A$, we notice that:
\begin{align}
    \label{eqn: action}
    U \ket{\bf 0}\ket{\phi} = \ket{\bf 0} A\ket{\phi} + \ket{\rm Garbage},
\end{align}
where $\ket{\rm Garbage }$ generally admits the form $\sum_{j \neq \bf 0} \ket{j}\ket{\rm Garbage}_j$ is a redundant state that is completely orthogonal to $\ket{\bf 0} A\ket{\phi}$, i.e, $\bra{\rm Garbage} \ket{\bf 0} A\ket{\phi} = 0$. 

\begin{lemma}[\cite{gilyen2019quantum}]
\label{lemma: improveddme}
Let $\rho = \Tr_A \ket{\Phi}\bra{\Phi}$, where $\rho \in \mathbb{H}_B$, $\ket{\Phi} \in  \mathbb{H}_A \otimes \mathbb{H}_B$. Given unitary $U$ that generates $\ket{\Phi}$ from $\ket{\bf 0}_A \otimes \ket{\bf 0}_B$, then there exists an efficient procedure that constructs an exact unitary block encoding of $\rho$.
\end{lemma}

The proof of the above lemma is given in \cite{gilyen2019quantum} (see their Lemma 45). \\

\begin{lemma}[Block Encoding of Product of Two Matrices]
\label{lemma: product}
    Given the unitary block encoding $U_1$, $U_2$ of two matrices $A_1$ and $A_2$ using $a$ and $b$ ancillas respectively, an efficient procedure exists that constructs a unitary block encoding of $A_1 A_2$. The total ancilla usage is $a+b$.
\end{lemma}

The proof of the above lemma is also given in~\cite{gilyen2019quantum}.  \\

\begin{lemma}[\cite{camps2020approximate}]
\label{lemma: tensorproduct}
    Given the unitary block encoding $\{U_i\}_{i=1}^m$ of multiple operators $\{M_i\}_{i=1}^m$ (assumed to be exact encoding), then, there is a procedure that produces the unitary block encoding operator of $\bigotimes_{i=1}^m M_i$, which requires a single use of each $\{U_i\}_{i=1}^m$ and $\mathcal{O}(1)$ SWAP gates. 
\end{lemma}
The above lemma is a result in~\cite{camps2020approximate}. 
\begin{lemma}
\label{lemma: As}
    Given the oracle access to $s$-sparse matrix $A$ of dimension $n\times n$, then an $\epsilon$-approximated unitary block encoding of $A/s$ can be prepared with gate/time complexity $\mathcal{O}(\log n + \log^{2.5}(\frac{s}{\epsilon}))$.
\end{lemma}
This is also presented in~\cite{gilyen2019quantum}.  One can also find similar construction in Ref.~\cite{childs2017lecture}. 
\begin{lemma}
    Given unitary block encoding $\{U_i\}_{i=1}^m$ of operators $\{M_i\}_{i=1}^m$. Then, there is a procedure that produces a unitary block encoding operator of $\sum_{i=1}^m (\gamma_i/\gamma) M_i $ in complexity $\mathcal{O}(m)$, where $\gamma = \sum_i \gamma_i$. The number of extra ancillas involved are $\mathcal{O}(  \log(m) + T_U  )$, where $T_U$ is the number of ancillas required to block encode $M_i$  for all $i$. 
    \label{lemma: sumencoding}
\end{lemma}

\begin{lemma}[Scaling Block encoding]
\label{lemma: scale}
    Given a block encoding of some matrix $A$ (as in~\ref{def: blockencode}), then the block encoding of $A/p$, where $p > 1$, can be prepared with an extra $\mathcal{O}(1)$ complexity. 
\end{lemma}

\begin{lemma}[\cite{gilyen2019quantum} Theorem 30]
\label{lemma: amp_amp}
Let $U$, $\Pi$, $\widetilde{\Pi} \in {\rm End}(\mathcal{H}_U)$ be linear operators on $\mathcal{H}_U$ such that $U$ is a unitary, and $\Pi$, $\widetilde{\Pi}$ are orthogonal projectors. 
Let $\gamma>1$ and $\delta,\epsilon \in (0,\frac{1}{2})$. 
Suppose that $\widetilde{\Pi}U\Pi=W \Sigma V^\dagger=\sum_{i}\varsigma_i\ket{w_i}\bra{v_i}$ is a singular value decomposition. 
Then there is an $m= \mathcal{O} \Big(\frac{\gamma}{\delta}
\log \left(\frac{\gamma}{\epsilon} \right)\Big)$ and an efficiently computable $\Phi\in\mathbb{R}^m$ such that
\begin{equation}
\left(\bra{+}\otimes\widetilde{\Pi}_{\leq\frac{1-\delta}{\gamma}}\right)U_\Phi \left(\ket{+}\otimes\Pi_{\leq\frac{1-\delta}{\gamma}}\right)=\sum_{i\colon\varsigma_i\leq \frac{1-\delta}{\gamma} }\tilde{\varsigma}_i\ket{w_i}\bra{v_i} , \text{ where } \Big|\!\Big|\frac{\tilde{\varsigma}_i}{\gamma\varsigma_i}-1 \Big|\!\Big|\leq \epsilon.
\end{equation}
Moreover, $U_\Phi$ can be implemented using a single ancilla qubit with $m$ uses of $U$ and $U^\dagger$, $m$ uses of C$_\Pi$NOT and $m$ uses of C$_{\widetilde{\Pi}}$NOT gates and $m$ single qubit gates.
Here,
\begin{itemize}
\item C$_\Pi$NOT$:=X \otimes \Pi + I \otimes (I - \Pi)$ and a similar definition for C$_{\widetilde{\Pi}}$NOT; see Definition 2 in \cite{gilyen2019quantum},
\item $U_\Phi$: alternating phase modulation sequence; see Definition 15 in \cite{gilyen2019quantum},
\item $\Pi_{\leq \delta}$, $\widetilde{\Pi}_{\leq \delta}$: singular value threshold projectors; see Definition 24 in \cite{gilyen2019quantum}.
\end{itemize}
\end{lemma}

\section{New Quantum Algorithm for Solving Linear Equations}
\label{sec: mainframework}
Recall that the main objective is to find $\xbf$ such that $A \xbf = \textbf{b}$, where $A$ is a Hermitian $n \times n$ matrix with sparsity $s$, and $\textbf{b}$ is a unit vector (assumed without loss of generalization). For simplicity, we work in the real regime. An additional assumption is that the magnitude of the eigenvalues of $A$ is between $(1/\kappa,1)$. As seen in the previous section, the direct solution is $\xbf = A^{-1} \textbf{b}$. An alternative, indirect solution to solving linear system is given by:
\begin{align}
\label{eqn: gradient}
    \min_{\xbf} \frac{1}{2} ||\xbf||^2 +  \frac{1}{2}|| A\xbf - \textbf{b} ||^2
\end{align}
where $||.||$ refers to $l_2$ Euclidean norm. Defining $f(\xbf) =  \frac{1}{2} ||\xbf||^2 + \frac{1}{2} || A\xbf - \textbf{b} ||^2$, a further expansion yields:
\begin{align}
    f(\xbf) &=  \frac{1}{2} ||\xbf||^2 \frac{1}{2} \left( ||A\xbf ||^2 -  \bra{\textbf{b}} A\xbf -  \xbf^\dagger A^\dagger \ket{\textbf{b}} + ||\textbf{b}||^2 \right) \\
       &=  \frac{1}{2} ||\xbf||^2 + \frac{1}{2}  ||A\xbf ||^2 -   \bra{\textbf{b}} A\xbf + \frac{1}{2} 
\end{align}
To find $\xbf$ so that $f(\xbf)$ is minimized, a standard/textbook method is gradient descent. The algorithm proceeds by first randomizing a point $\xbf_0$, then iterate the following procedure:
\begin{align}
    \xbf_{t+1} = \xbf_t - \eta \bigtriangledown f(\xbf_t)
\end{align}
where $\xbf_t$ refers to the solution at $t$-th iteration, and $\eta$ is the hyperparameter. In general, the total number of iterations $T$ is typically user-dependent, and the convergence of the gradient descent algorithm depends on the behavior of the function. It turns out that the function $f(\xbf)$ above is strongly convex, which means that the convergence of gradient descent algorithm is guaranteed. In particular, Ref.~\cite{nesterov2013introductory, nesterov1983method, boyd2004convex, garrigos2023handbook} proved that choosing $T = \mathcal{O}\big( \frac{1}{\eta} \log \frac{1}{\delta}\big)$ suffices to secure that $x_T$ is $\delta$-close to the true solution. 

We remark that a crucial part of gradient descent method is the evaluation of the given function's gradient $\bigtriangledown f(\xbf)$. In theory, if one can compute the function $f(\xbf) \equiv f(x_1,x_2,...,x_n)$ at any point within the working domain, then the partial derivative of $f$, for example, with respect to $x_1$, can be numerically computed as:
\begin{align}
    \frac{\partial f(x_1,x_2,...,x_n)}{\partial x_1} \approx \frac{f(x_1+\delta,x_2,...,x_n)- f(x_1,x_2,...,x_n)}{\delta}
\end{align}
Similarly, the partial derivative with respect to remaining variables can be numerically computed. So, the gradient $\bigtriangledown f(\xbf) = \Big( \frac{\partial f(\xbf)}{\partial x_1},  \frac{\partial f(\xbf)}{\partial x_2},...,  \frac{\partial f(\xbf)}{\partial x_n}  \Big)^T$ can be constructed. We note that the first-order derivative approximation formula provided above is the simplest one, and there are more complicated yet more accurate ones. In principle, given oracle access to $A$, a means to prepare $\ket{\textbf{b}}$ and suppose that we also have a means to prepare $\xbf$, the function $f(\xbf)$ can be estimated. A concrete procedure will be provided in Appendix~\ref{sec: evaluationfunctionf}. Here, we point out that if we use such an approach to estimate the function $f$ at some $\xbf$ plus nearby points $\xbf + \Vec{\delta}$, and use that information to estimate the entries of $\bigtriangledown f(\xbf)$, then constructing the state $\sim \bigtriangledown f(\xbf)$ can be done by amplitude encoding \cite{grover2000synthesis,grover2002creating,plesch2011quantum, schuld2018supervised, nakaji2022approximate,marin2023quantum,zoufal2019quantum, zhang2022quantum}. However, as the vector $\bigtriangledown f(\xbf)$ is not sparse, it can be costly, hence prohibiting potential speedup in dimension $n$. Additionally, the error from estimation of the gradient' entries accumulates, resulting in an overall error of order $n$ in the gradient $\bigtriangledown f(\xbf)$. 

It turns out that the function $f(\xbf)$ admits an analytical form for the gradient as following:
\begin{align}
\label{eqn: analyticalgradient}
    \bigtriangledown f(\xbf) &= \xbf +  \big(  A^\dagger A \big)  \xbf - A^\dagger \ket{\textbf{b}}  \\
    &= \big( \Ibb + A^\dagger A \big) \xbf - A^\dagger \ket{\textbf{b}} 
\end{align}
We remark that a technical aspect that distinguishes our work to existing ones, such as \cite{harrow2009quantum, childs2017quantum, kerenidis2020quantum,huang2019near, subacsi2019quantum}, is that instead of vector/state representation, we use density operator representation. More concretely, instead of $\xbf$, we use $\xbf \xbf^\dagger$ as a main working object. This choice allows us to leverage the block-encoding and related techniques mentioned in the previous section. Subsequently, we will show that from density operator, we can recover the quantum state corresponding to the solution of the linear systems. 

In the new representation, the gradient descent updating step is:
\begin{align}
\label{eqn: gradientstep}
    \Big( \xbf - \eta \bigtriangledown f(\xbf)  \Big) \Big( \xbf - \eta \bigtriangledown f(\xbf)  \Big)^\dagger &= \xbf \xbf^\dagger - \eta \xbf \bigtriangledown f(\xbf)^\dagger - \eta \bigtriangledown f(\xbf) \xbf^\dagger + \eta^2 \bigtriangledown f(\xbf) \bigtriangledown f^\dagger (\xbf) \\
    &= \xbf \xbf^\dagger -  \eta \xbf  \Big(\big( \Ibb + A^\dagger A \big)  \xbf -  A^\dagger \ket{\textbf{b}}  \Big)^\dagger - \eta \Big(  \big( \Ibb + A^\dagger A \big)  \xbf -  A^\dagger \ket{\textbf{b}}  \Big)\xbf^\dagger  \\ &+ \eta^2 \Big( \big( \Ibb + A^\dagger A \big)  \xbf -  A^\dagger \ket{\textbf{b}}  \Big) \Big( \big( \Ibb + A^\dagger A \big)  \xbf -   A^\dagger \ket{\textbf{b}}  \Big)^\dagger \\ 
\end{align}
We expand term by term as follows.
\begin{align}
    \eta \xbf  \Big(\big( \Ibb + A^\dagger A \big)  \xbf -   A^\dagger \ket{\textbf{b}}  \Big)^\dagger &= \eta \left( \xbf \xbf^\dagger \big( \Ibb + A^\dagger A \big)  - \xbf  \bra{\textbf{b}} A \right) \\
    \eta  \Big(  \big( \Ibb + A^\dagger A \big)  \xbf -   A^\dagger \ket{\textbf{b}} \Big)\xbf^\dagger  &= \eta \left( \big( \Ibb + A^\dagger A \big)  \xbf \xbf^\dagger -  A^\dagger \ket{\textbf{b}} \xbf^\dagger \right) \\
    \eta^2 \Big( \big( \Ibb + A^\dagger A \big)  \xbf -  A^\dagger \ket{\textbf{b}}  \Big) \Big( \big( \Ibb + A^\dagger A \big)  \xbf -  A^\dagger \ket{\textbf{b}}  \Big)^\dagger &= \eta^2 \Big( \big( \Ibb + A^\dagger A \big)  \xbf \xbf^\dagger \big( \Ibb + A^\dagger A \big)   - \big( \Ibb + A^\dagger A \big)  \xbf \bra{\bbf} A \\ & -   A^\dagger \ket{\bbf} \xbf^\dagger \big( \Ibb + A^\dagger A \big)  +  A^\dagger \ket{\bbf}\bra{\bbf}A \Big)
\end{align}
Our algorithm makes use of the following: \\

(1) Suppose there is a unitary $U_b$ that generates $\ket{\bbf}$, then Lemma \ref{lemma: improveddme} allows us to block encode the operator $\ket{\bbf}\bra{\bbf}$ \\

(2) Oracle access to entries of $A$ can be combined with Lemma \ref{lemma: As} to construct an $\epsilon$-closed block encoding of $A/s$, denoted as $U_A$. Then $U_A^\dagger$ is the block encoding of $A^\dagger/s$. Use Lemma \ref{lemma: product} to construct the block encoding of $ \frac{1}{s^2}A^\dagger A$. Given that the block encoding of $\Ibb$ is trivial (see discussion below Definition \ref{def: blockencode}), Lemma \ref{lemma: sumencoding} can be used to construct the block encoding of $\frac{1}{2s^2} \Big( \Ibb + A^\dagger A \Big)$. \\

(3) Suppose for now that we have a block encoding of operator $\alpha \xbf \xbf^\dagger$ (where $\alpha$ is some factor). Then the inner product $\alpha \xbf^\dagger \ket{\bbf}$ (including sign and magnitude) can be estimated to a known precision. A concrete procedure is provided in Appendix \ref{sec: evaluationinnerpdouct}.  \\

\noindent
\textbf{Quantum Algorithm For Solving Linear Systems of Equations.} We proceed to describe our new proposal for solving linear systems of equations as follows. The aim is to construct the block encoding of operators in Eqn.~\ref{eqn: mainoperators}. First, we use a random unitary $U_0$ to generate an arbitrary state $\ket{\phi}$ (of dimension $> n$) that contains $\xbf_0$ as its first $n$ entries. Using Lemma \ref{lemma: improveddme}, we can construct a block encoding of $\ket{\phi}\bra{\phi}$. The top-left corner of such a matrix is $\xbf_0 \xbf_0^\dagger$. According to Def.~\ref{def: blockencode}, it is also a block encoding of $\xbf_0\xbf_0^\dagger$. \\ 

\noindent
\textit{Step 1: Constructing block encoding of  $\alpha \frac{\xbf_0^\dagger \ket{\bbf}}{8} \Big( \xbf_0 \xbf_0^\dagger (\Ibb + A^\dagger A) - \xbf \bra{\bbf} A \Big) $ ( $0<\alpha <1$ is some factor will be used to define hyperparameter $\eta$)}
\begin{itemize}
    \item First use (3) (also result of Appendix \ref{sec: evaluationinnerpdouct}) to estimate $ \xbf_0^\dagger \ket{\bbf}$ to a known accuracy, namely $\epsilon$. 
    
    \item The block encoding of $\xbf_0\xbf_0^\dagger$ and of $\ket{\bbf}\bra{\bbf}$ can be used with Lemma \ref{lemma: product} to obtain the block encoding of $\xbf_0\xbf_0^\dagger \ \ket{\bbf}\bra{\bbf} = \xbf_0^\dagger \ket{\bbf} \xbf_0 \bra{\bbf}$.
    
    \item Use the block encoding of $\xbf_0\xbf_0^\dagger$, $\ket{\bbf}\bra{\bbf}$ and of $A/s$ to construct the block encoding of $ \xbf_0 \xbf_0^\dagger \ \ket{\bbf}\bra{\bbf} \frac{A}{s} = \frac{\xbf_0^\dagger}{s} \ket{\bbf}  \xbf_0\bra{\bbf} A$
    
    \item Apply Lemma \ref{lemma: scale} (with scaling factor $p = 2s$) to $ \frac{\xbf_0^\dagger \ket{\bbf} }{s} \xbf_0\bra{\bbf} A $ to transform it into the block encoding of $  \frac{\xbf_0^\dagger \ket{\bbf} }{2s^2} \xbf_0\bra{\bbf} A $. 
    
    \item Use Lemma \ref{lemma: product} with the block encoding of $\xbf_0 \xbf_0^\dagger$ and of $A/s, A^\dagger/s$ to construct the unitary block encoding of $\xbf_0 \xbf_0^\dagger \cdot \big( \Ibb + A^\dagger A \big) /(2s^2)$. 

    \item Apply Lemma \ref{lemma: scale} (with scaling factor $p = 1/(\xbf_0^\dagger \ket{\bbf})$) to the block encoding of $\frac{1}{2s^2}\xbf_0\xbf_0^\dagger \big( \Ibb + A^\dagger A \big)   $ to transform it into the block encoding of $\frac{\xbf_0^\dagger \ket{\bbf} }{2s^2}\xbf_0 \xbf_0^\dagger \big( \Ibb + A^\dagger A \big)   $

    \item Use Lemma \ref{lemma: sumencoding} to construct the block encoding of:
    \begin{align}
        \frac{1}{2} \Big( \frac{\xbf_0^\dagger \ket{\bbf} }{2s^2}\xbf_0 \xbf_0^\dagger \big( \Ibb + A^\dagger A \big)   - \frac{\xbf_0^\dagger \ket{\bbf} }{2s^2} \xbf_0\bra{\bbf} A  \Big) &= \frac{\xbf_0^\dagger \ket{\bbf} }{4 s^2 } \Big(  \xbf_0 \xbf_0^\dagger \big( \Ibb + A^\dagger A \big)  - \xbf_0\bra{\bbf} A \Big)  \\
        &= \frac{\xbf_0^\dagger \ket{\bbf} }{4 s^2 }  \xbf_0 \Big( \big( \Ibb + A^\dagger A \big) \xbf_0 - A^\dagger \ket{\bbf}  \Big)^\dagger
    \end{align}
    \item Use Lemma \ref{lemma: scale} with $p=1/\alpha$ to transform the block-encoded operator above into $\alpha\frac{ \xbf_0^\dagger \ket{\bbf} }{4 s^2 }  \xbf_0 \Big( \big( \Ibb + A^\dagger A \big) \xbf_0 - A^\dagger \ket{\bbf}  \Big)^\dagger $ 
    \item Use Lemma \ref{lemma: amp_amp} to remove the factor $s^2$, e.g., apply Lemma \ref{lemma: amp_amp} with $\gamma= s^2/2$ to the above block-encoded operator, we obtain the block encoding of $ \frac{\xbf_0^\dagger \ket{\bbf} }{8 }  \xbf_0 \Big( \big( \Ibb + A^\dagger A \big) \xbf_0 - A^\dagger \ket{\bbf}  \Big)^\dagger$ 
\end{itemize}

\noindent
\textit{Step 2: Constructing block encoding of $\alpha \frac{\bra{\bbf}\xbf_0}{8}\Big( (\Ibb+A^\dagger A) - A^\dagger \ket{\bbf}\xbf^\dagger \Big)$}
\begin{itemize}
    \item The transpose of the above unitary block encoding is the block encoding of
    \begin{align}
        \alpha  \frac{  \bra{\bbf}\xbf_0 }{8} \Big( \big( \Ibb + A^\dagger A \big)  \xbf_0 \xbf_0^\dagger  - A^\dagger \ket{\bbf}\xbf_0^\dagger  \Big) = \alpha \frac{ \bra{\bbf}\xbf_0}{8 }  \Big(  (\Ibb+AA^\dagger) \xbf_0 - A^\dagger \ket{\bbf} \Big) \xbf_0^\dagger 
    \end{align}
\end{itemize}

\noindent
\textit{Step 3: Constructing block encoding of $\alpha^2 \frac{\xbf_0^\dagger \ket{\bbf  }}{64}\Big( \big( \Ibb + A^\dagger A \big)  \xbf_0 \xbf_0^\dagger \big( \Ibb + A^\dagger A \big)   -\big( \Ibb + A^\dagger A \big)  \xbf_0\bra{\bbf} A  -   A^\dagger \ket{\bbf} \xbf_0^\dagger \big( \Ibb + A^\dagger A \big)  \\ +  A^\dagger \ket{\bbf}\bra{\bbf}A  \Big)$}
\begin{itemize}
    \item Use the block encoding of $\xbf_0^\dagger \ket{\bbf}  \xbf_0\bra{\bbf} A/s, \bra{\bbf} \xbf_0 (A^\dagger/s) \ket{\bbf}\xbf_0^\dagger  $ and of $\frac{1}{2s}(\Ibb + A^\dagger A)$ to construct the block encoding of
    \begin{align}
         \frac{\big( \Ibb + A^\dagger A \big)  }{2s^2} \xbf_0^\dagger \ket{\bbf}  \xbf_0\bra{\bbf}  \frac{A}{s}, \  \bra{\bbf} \xbf_0 \frac{A^\dagger}{s}\ket{\bbf}\xbf_0^\dagger  \frac{\big( \Ibb + A^\dagger A \big) }{2s^2}       
    \end{align}
    and use Lemma \ref{lemma: scale} with $p = 2s$ to multiply the above block-encoded operator with $1/2s$, obtaining:
     \begin{align}
         \frac{\big( \Ibb + A^\dagger A \big)  }{4s^4} \xbf_0^\dagger \ket{\bbf}  \xbf_0\bra{\bbf}  A, \  \bra{\bbf} \xbf_0 \frac{A^\dagger}{4s^4}\ket{\bbf}\xbf_0^\dagger  \big( \Ibb + A^\dagger A \big)      
    \end{align}
    \item Similarly, we can obtain the block encoding of $ \frac{1}{4s^4} \big( \Ibb + A^\dagger A \big)  \xbf_0 \xbf_0^\dagger \big( \Ibb + A^\dagger A \big)  $. 
    \item Use Lemma \ref{lemma: scale} with scaling factor $p =1/\braket{\xbf_0,b}$ to construct the block encoding of 
    \begin{align}
        \frac{\braket{\xbf_0,b}}{4s^4 } \big( \Ibb + A^\dagger A \big)  \xbf_0 \xbf_0^\dagger \big( \Ibb + A^\dagger A \big)  
    \end{align}
    \item Use block encoding of $\ket{\bbf}\bra{\bbf} $ and of $A/s,A^\dagger/s$ with Lemma \ref{lemma: product} to construct the block encoding of 
    \begin{align}
        \frac{1}{s^2} A^\dagger \ket{\bbf}\bra{\bbf} A
    \end{align}
    then use Lemma \ref{lemma: scale} with $p= s^2/ \xbf_0^\dagger \ket{\bbf} $ to transform the above operator into 
    \begin{align}
        \frac{\xbf_0^\dagger \ket{\bbf} }{s^4 } A^\dagger \ket{\bbf}\bra{\bbf} A
    \end{align}
    \item Use Lemma \ref{lemma: sumencoding} to construct the block encoding of:
    \begin{align}
      &\frac{1}{4}\left(  \frac{\braket{\xbf_0,b}}{4s^4 } \big( \Ibb + A^\dagger A \big)  \xbf_0 \xbf_0^\dagger \big( \Ibb + A^\dagger A \big)  - \frac{\big( \Ibb + A^\dagger A \big)  }{4s^4} \xbf_0^\dagger \ket{\bbf}  \xbf_0\bra{\bbf}  A-  \bra{\bbf} \xbf_0 \frac{A^\dagger}{4s^4}\ket{\bbf}\xbf_0^\dagger  \big( \Ibb + A^\dagger A \big)     + \frac{\xbf_0^\dagger \ket{\bbf} }{4s^4 } A^\dagger \ket{\bbf}\bra{\bbf} A \right) \\
      &= \frac{\xbf_0^\dagger \ket{\bbf} }{16 s^4} \Big( \big( \Ibb + A^\dagger A \big)  \xbf_0 \xbf_0^\dagger \big( \Ibb + A^\dagger A \big)  - \big( \Ibb + A^\dagger A \big)  \xbf_0\bra{\bbf} A - A^\dagger \ket{\bbf}\xbf_0^\dagger \big( \Ibb + A^\dagger A \big)  + A^\dagger \ket{\bbf }\bra{\bbf} A \Big)  \\
      &= \frac{\xbf_0^\dagger \ket{\bbf} }{16 s^4} \Big( \big( \Ibb + A^\dagger A \big)  \xbf_0 - A^\dagger\ket{\bbf}  \Big)\Big(  \big( \Ibb + A^\dagger A \big)  \xbf_0 - A^\dagger \ket{\bbf} \Big)^\dagger
    \end{align}
    \item Use Lemma \ref{lemma: scale} (with $p=1/\alpha^2$) to transform the above operator into $\alpha^2\frac{\xbf_0^\dagger \ket{\bbf} }{16 s^4} \Big( \big( \Ibb + A^\dagger A \big)  \xbf_0 - A^\dagger\ket{\bbf}  \Big)\Big(  \big( \Ibb + A^\dagger A \big)  \xbf_0 - A^\dagger \ket{\bbf} \Big)^\dagger $
    \item Use Lemma \ref{lemma: amp_amp} with $\gamma = s^4/4$ to the above operator, we obtain the block encoding of $\alpha^2\frac{\xbf_0^\dagger \ket{\bbf} }{64} \Big( \big( \Ibb + A^\dagger A \big)  \xbf_0 - A^\dagger\ket{\bbf}  \Big)\Big(  \big( \Ibb + A^\dagger A \big)  \xbf_0 - A^\dagger \ket{\bbf} \Big)^\dagger $
    \end{itemize}

\noindent
\textit{Step 4: Combining altogether.}
    \begin{itemize}
    \item For convenience, recall that we have block encoding of the following operators:
    \begin{align}
    \label{eqn: mainoperators}
        \xbf_0 \xbf_0^\dagger \\ 
       \alpha \frac{\xbf_0^\dagger \ket{\bbf} }{4 s^2 }  \xbf_0 \Big( \big( \Ibb + A^\dagger A \big) \xbf_0 - A^\dagger \ket{\bbf}  \Big)^\dagger & = \alpha \frac{\xbf_0^\dagger \ket{\bbf} }{8 }  \xbf_0 \bigtriangledown f^\dagger (\xbf_0)  \\
        \alpha \frac{ \bra{\bbf}\xbf_0}{4 s^2 }  \Big(  (\Ibb +A^\dagger A) \xbf_0 - A^\dagger \ket{\bbf} \Big) \xbf_0^\dagger   & = \alpha \frac{ \bra{\bbf}\xbf_0}{8} \bigtriangledown f(\xbf_0) \xbf_0^\dagger  \\
       \alpha^2 \frac{\xbf_0^\dagger \ket{\bbf} }{16 s^4} \Big( \big( \Ibb + A^\dagger A \big)  \xbf_0 - A^\dagger\ket{\bbf}  \Big)\Big(  \big( \Ibb + A^\dagger A \big)  \xbf_0 - A^\dagger \ket{\bbf} \Big)^\dagger &=\alpha^2 \frac{\xbf_0^\dagger \ket{\bbf} }{64} \bigtriangledown f(\xbf_0 ) \bigtriangledown  f^\dagger(\xbf_0)
    \end{align}
    where for convenience, we remind from Eqn. \ref{eqn: analyticalgradient} that:
    \begin{align}
        \bigtriangledown f(\xbf) = \big( \Ibb + A^\dagger A \big)  \xbf - A^\dagger \ket{\textbf{b}} 
    \end{align}
    \item Use Lemma \ref{lemma: scale} with $p = 1/(\xbf_0^\dagger \ket{\bbf} )$ to transform the block-encoded operator $\xbf_0 \xbf_0^\dagger \longrightarrow \xbf_0^\dagger \ket{\bbf} \xbf_0 \xbf_0^\dagger $
    \item Use Lemma \ref{lemma: sumencoding} to construct the block encoding of:
    \begin{align}
    &\frac{1}{4}\Big( \xbf_0^\dagger \ket{\bbf}  \xbf_0 \xbf_0^\dagger - \alpha \frac{\xbf_0^\dagger \ket{\bbf} }{8} \xbf_0\bigtriangledown f^\dagger (\xbf_0) - \alpha \frac{ \bra{\bbf}\xbf_0}{8} \bigtriangledown f(\xbf_0) \xbf_0^\dagger  +  \alpha ^2\frac{\xbf_0^\dagger \ket{\bbf} }{64} \bigtriangledown f(\xbf_0 ) \bigtriangledown f^\dagger(\xbf_0)  \Big)  \\
    &= \frac{\xbf_0^\dagger \ket{\bbf} }{4} \Big(   \xbf_0 \xbf_0^\dagger -  \frac{\alpha }{8 }\xbf_0\bigtriangledown f^\dagger (\xbf_0) - \frac{\alpha }{8} \bigtriangledown f(\xbf_0) \xbf_0^\dagger + \frac{\alpha^2}{64}\bigtriangledown f(\xbf_0 ) \bigtriangledown f^\dagger(\xbf_0) \Big) 
    \end{align}
    \item Define the hyperparameter $\eta = \frac{\alpha  }{8}$. According to Eqn.~\ref{eqn: gradientstep}, the above equation is:
    \begin{align}
        & \frac{\xbf_0^\dagger \ket{\bbf} }{4} \Big(   \xbf_0 \xbf_0^\dagger -  \eta \xbf_0\bigtriangledown f^\dagger (\xbf_0) - \eta \bigtriangledown f(\xbf_0) \xbf_0^\dagger + \eta^2 \bigtriangledown f(\xbf_0 ) \bigtriangledown f^\dagger(\xbf_0) \Big)  \\
        &=\frac{\xbf_0^\dagger \ket{\bbf} }{4} \Big( \xbf_0 - \eta \bigtriangledown f(\xbf_0) \Big) \Big( \xbf_0 - \eta \bigtriangledown f(\xbf_0)  \Big)^\dagger \\&= \frac{\xbf_0^\dagger \ket{\bbf} }{4} \xbf_1 \xbf_1^\dagger
    \end{align}
    where $\xbf_1 = \xbf_0 - \eta \bigtriangledown f(\xbf_0)$ is the temporal solution after the first gradient descent step. 
    \item Use the block encoding of the above operator $ \frac{\xbf_0^\dagger \ket{\bbf}  }{4} \xbf_1 \xbf_1^\dagger$ as an input, and repeat from the beginning, e.g., Step 1 - Step 4. Note that there is a factor $ \xbf_0^\dagger \ket{\bbf} /4$, so the outcome of the next iteration is gonna be:
    \begin{align}
        \frac{\big(\xbf_0^\dagger \ket{\bbf}\big) }{4} \frac{\big(\xbf_1^\dagger\ket{\bbf}\big)}{4} \Big( \xbf_1 - \eta \bigtriangledown f(\xbf_1)   \Big) \Big( \xbf_1 - \eta \bigtriangledown f(\xbf_1) \Big)^\dagger = \frac{ \big(\xbf_0^\dagger \ket{\bbf}\big) }{4} \frac{ \big(\xbf_1^\dagger\ket{\bbf} \big)}{4}  \xbf_2 \xbf_2^\dagger
    \end{align}
    where again $\xbf_2 = \xbf_1 - \eta \bigtriangledown f(\xbf_1)$. Continually, we obtain the block encoding of:
    \begin{align}
        \frac{ \xbf_0^\dagger \ket{\bbf} \xbf_1^\dagger \ket{\bbf} \xbf_2^\dagger \ket{\bbf}}{4^3} \xbf_3 \xbf_3^\dagger, \frac{ \xbf_0^\dagger \ket{\bbf} \xbf_1^\dagger \ket{\bbf} \xbf_2^\dagger \ket{\bbf} \xbf_3^\dagger \ket{\bbf}}{4^4} \xbf_4\xbf_4^\dagger, ..., \frac{ \xbf_0^\dagger \ket{\bbf} \xbf_1^\dagger \ket{\bbf} \xbf_2^\dagger \ket{\bbf}... \xbf_{T-1}^\dagger \ket{\bbf} }{4^T} \xbf_T \xbf_T^\dagger
    \end{align}
    \item Let $U_T$ be the unitary block encoding of $\frac{ \xbf_0^\dagger \ket{\bbf} \xbf_1^\dagger \ket{\bbf} \xbf_2^\dagger \ket{\bbf}... \xbf_{T-1}^\dagger \ket{\bbf} }{4^T} \xbf_T \xbf_T^\dagger $. According to Definition \ref{def: blockencode} (and also related property Eqn.~\ref{eqn: action}), we have:
    \begin{align}
        U_T \ket{\bf 0}\ket{\bbf} =  \ket{\bf 0} \Big( \frac{ \xbf_0^\dagger \ket{\bbf} \xbf_1^\dagger \ket{\bbf} \xbf_2^\dagger \ket{\bbf}... \xbf_{T-1}^\dagger \ket{\bbf} }{4^T}  \Big) \xbf_T \xbf_T^\dagger \ket{\bbf} + \ket{\rm Garbage}
    \end{align}
    Measuring the ancilla register and post-select $\ket{\bf 0}$, we obtain the state $\ket{\xbf_T} = \xbf_T/||\xbf_T||$, which is the solution to the optimization problem (Eqn.\ref{eqn: gradient}), and also of the linear system $A\xbf = \bbf$
\end{itemize}

\section{Analysis and Discussion}
\label{sec: analysis}
In the following, we provide an analysis and discuss some aspects of the above algorithm. \\

\noindent
\textbf{Error.} First, we analyze the error. Considering at $t$-th time step, we have some $\epsilon_t$-approximated block encoding of
\begin{align}
   \textbf{X}_t \equiv  \frac{ \xbf_0^\dagger \ket{\bbf} \xbf_1^\dagger \ket{\bbf} \xbf_2^\dagger \ket{\bbf}... \xbf_{t-1}^\dagger \ket{\bbf} }{4^t} \xbf_t \xbf_t^\dagger 
\end{align}
The first step is to produce the block encoding of the following operator (Eqn.~\ref{eqn: mainoperators}):
\begin{align}
    & \frac{ \xbf_0^\dagger \ket{\bbf} \xbf_1^\dagger \ket{\bbf} \xbf_2^\dagger \ket{\bbf}... \xbf_{t-1}^\dagger \ket{\bbf} \xbf_t^\dagger \ket{\bbf} }{4^t}  \eta \xbf_t \bigtriangledown  f^\dagger (\xbf_t) \\
    & \frac{ \xbf_0^\dagger \ket{\bbf} \xbf_1^\dagger \ket{\bbf} \xbf_2^\dagger \ket{\bbf}... \xbf_{t-1}^\dagger \ket{\bbf} \xbf_t^\dagger \ket{\bbf} }{4^t} \frac{\alpha }{8} \Big( \xbf_t \xbf_t^\dagger \big( \Ibb + A^\dagger A \big)  - \xbf_t \bra{\bbf} A \Big)  \\
     &=  \frac{\alpha\xbf_t^\dagger \ket{\bbf}}{8} \textbf{X}_t  \big( \Ibb + A^\dagger A \big)  -   \frac{\alpha}{8} \textbf{X}_t  \ket{\bbf} \bra{\bbf} A
\end{align}
As $A/s,A^\dagger/s$ is each block-encoded with accuracy $\epsilon$, the error adds up linearly, the total error is $ \epsilon_t + 2\epsilon + \epsilon_t + \epsilon = 2\epsilon_t + 3\epsilon$. It means that we obtain an $(2\epsilon_t+3\epsilon)$-approximated block encoding of the above operator. Similarly, also from Eqn.~\ref{eqn: mainoperators}, we obtain $(2\epsilon_t + 3\epsilon)$-approximated block encoding of the transpose of the above operator, i.e,.
\begin{align}
    \frac{\Big( \xbf_0^\dagger \ket{\bbf} \xbf_1^\dagger \ket{\bbf} \xbf_2^\dagger \ket{\bbf}... \xbf_{t-1}^\dagger \ket{\bbf} \xbf_t^\dagger \ket{\bbf} \Big)^\dagger}{4^t}  \eta \bigtriangledown f (\xbf_t)\xbf_t^\dagger
\end{align}
The last is the production of:
\begin{align}
&= \frac{ \xbf_0^\dagger \ket{\bbf} \xbf_1^\dagger \ket{\bbf} \xbf_2^\dagger \ket{\bbf}... \xbf_{t-1}^\dagger \ket{\bbf} \xbf_t^\dagger \ket{\bbf} }{4^t} \eta^2 \bigtriangledown f(\xbf_t) \bigtriangledown f^\dagger (\xbf_t)  \\
& \frac{ \xbf_0^\dagger \ket{\bbf} \xbf_1^\dagger \ket{\bbf} \xbf_2^\dagger \ket{\bbf}... \xbf_{t-1}^\dagger \ket{\bbf} \xbf_t^\dagger \ket{\bbf} }{4^t} \frac{\alpha^2}{64}\Big( \big( \Ibb + A^\dagger A \big)  \xbf_t \xbf_t^\dagger \big( \Ibb + A^\dagger A \big)  - \big( \Ibb + A^\dagger A \big)  \xbf_t \bra{\bbf} A - A^\dagger \ket{\bbf} \xbf_t^\dagger \big( \Ibb + A^\dagger A \big)  + A^\dagger \ket{\bbf}\bra{\bbf}A  \Big) \\
&= \xbf_t^\dagger \ket{\bbf}\frac{\alpha^2}{64}  \big( \Ibb + A^\dagger A \big)  \textbf{X}_t \big( \Ibb + A^\dagger A \big)  -\frac{\alpha^2}{64} \big( \Ibb + A^\dagger A \big)  \textbf{X}_t \ket{\bbf}\bra{\bbf} A - \frac{\alpha^2}{64} A^\dagger \ket{\bbf}\bra{\bbf} \textbf{X}_t \big( \Ibb + A^\dagger A \big)  +   \frac{\alpha^2\xbf_t^\dagger \ket{\bbf}}{64}  A^\dagger \ket{\bbf}\bra{\bbf} A
\end{align}
Using the same reasoning, the total error accumulated is $\epsilon_t + 4\epsilon + \epsilon_t + 3\epsilon + \epsilon_t + 3\epsilon + \epsilon_t+ 3\epsilon =  4\epsilon_t + 13\epsilon  $. The final step is producing the block encoding of:
\begin{align}  
    &\frac{ \xbf_0^\dagger \ket{\bbf} \xbf_1^\dagger \ket{\bbf} \xbf_2^\dagger \ket{\bbf}... \xbf_{t-1}^\dagger \ket{\bbf} \xbf_t^\dagger \ket{\bbf} }{4^{t}} \frac{1}{4}\Big( \xbf_t\xbf_t^\dagger - \eta \xbf_t \bigtriangledown f^\dagger (\xbf_t) - \eta \bigtriangledown f(\xbf_t) \xbf_t^\dagger + \eta^2 \bigtriangledown f(\xbf_t)\bigtriangledown f^\dagger (\xbf_t)   \Big)  
\end{align}
The error induced in producing a block encoding of $\sim \xbf_{t+1}\xbf_{t+1}^\dagger$ is $\epsilon_{t+1} = \frac{1}{4} \Big( \epsilon_t + 2\epsilon_t + 3\epsilon + 2\epsilon_t + 3\epsilon + 4\epsilon_t + 13\epsilon \Big) = \frac{1}{4}\Big(  9\epsilon_t + 19\epsilon \Big)$. By induction, we have that $\epsilon_t = \frac{1}{4}\Big( 9\epsilon_{t-1} + 19\epsilon  \Big)$. So we have:
\begin{align}
    \epsilon_{t+1} = \Big( \frac{9}{4}\Big)^2 \epsilon_{t-1} + \frac{9}{4} \frac{19}{4} \epsilon + \frac{19}{4}\epsilon
\end{align}
Similarly, by substituting $\epsilon_{t-1} =\frac{9}{4}\epsilon_{t-2} + \frac{19}{4}\epsilon$, we have:
\begin{align}
    \epsilon_{t+1} &= \Big( \frac{9}{4}\Big)^3 \epsilon_{t-2} + \Big( \frac{9}{4} \Big)^2 \frac{19}{4} \epsilon + \frac{9}{4} \frac{19}{4} \epsilon + \frac{19}{4}\epsilon \\
    &= \Big( \frac{9}{4}\Big)^4 \epsilon_{t-3} + \Big( \frac{9}{4} \Big)^3 \frac{19}{4} \epsilon 
 +  \Big( \frac{9}{4} \Big)^2 \frac{19}{4} \epsilon + \frac{9}{4} \frac{19}{4} \epsilon + \frac{19}{4}\epsilon \\
  & \vdots \\
  & = \Big( \frac{9}{4}\Big)^{t} \epsilon_{t=1} + \frac{19}{4}\epsilon \sum_{k=0}^{t-1} \Big( \frac{9}{4}\Big)^k \\
  &= \Big( \frac{9}{4}\Big)^{t} \epsilon_{t=1} + \frac{19}{4}\epsilon \frac{(9/4)^t - 1}{(9/4)-1}
\end{align}
Set $\epsilon_{t=1} = \epsilon$. Thus for a total of $T$ iteration, we have: 
\begin{align}
    \epsilon_T = \mathcal{O}\Big( \big(\frac{9}{4}\big)^{T-1} \epsilon  \Big)
\end{align}
In order to obtain the overall error of $\delta$, which means that we obtain the $\delta$-approximated block encoding of 
$$ \frac{ \xbf_0^\dagger \ket{\bbf} \xbf_1^\dagger \ket{\bbf} \xbf_2^\dagger \ket{\bbf}... \xbf_{T-1}^\dagger \ket{\bbf} }{4^T} \xbf_T \xbf_T^\dagger $$
we need to set $\epsilon_T= \delta$ and thus $ \epsilon = \mathcal{O}\Big( \delta \big( \frac{4}{9}\big)^{T-1} \Big)$. \\

\noindent
\textbf{Complexity.} Second, we analyze the total complexity for obtaining $U_T$, which is the unitary block encoding of
$$ \frac{ \xbf_0^\dagger \ket{\bbf} \xbf_1^\dagger \ket{\bbf} \xbf_2^\dagger \ket{\bbf}... \xbf_{T-1}^\dagger \ket{\bbf} }{4^T} \xbf_T \xbf_T^\dagger $$
As there are totally $T$ iteration steps, we use induction. Let $\mathscr{C}(\xbf_t)$ refer to the complexity required to produce the block encoding of
\begin{align}
   \textbf{X}_t \equiv  \frac{ \xbf_0^\dagger \ket{\bbf} \xbf_1^\dagger \ket{\bbf} \xbf_2^\dagger \ket{\bbf}... \xbf_{t-1}^\dagger \ket{\bbf} }{4^t} \xbf_t \xbf_t^\dagger
\end{align}
Evidently, it takes two block encodings of the above operator and three block encodings of $A/s,A^\dagger/s$ to produce the block encoding of 
\begin{align}
  \frac{ \xbf_0^\dagger \ket{\bbf} \xbf_1^\dagger \ket{\bbf} \xbf_2^\dagger \ket{\bbf}... \xbf_{t-1}^\dagger \ket{\bbf} \xbf_t^\dagger \ket{\bbf} }{4^t} \frac{\alpha}{4s^2} \Big(   \xbf_t \xbf_t^\dagger \big( \Ibb + A^\dagger A \big)  -  \xbf_t \bra{\bbf} A \Big) 
\end{align}
The application of Lemma \ref{lemma: amp_amp} to remove the factor $s^2$ takes further $\mathcal{O}(s^2)$ uses of block encoding of the above operator. So in total there is $2 \mathcal{O}(s^2)$ uses of block encoding of $\textbf{X}_t$ and $3 \mathcal{O}(s^2)$ uses of block encoding of $A/s,A^\dagger/s$ for producing block encoding of:
\begin{align}
    \frac{ \xbf_0^\dagger \ket{\bbf} \xbf_1^\dagger \ket{\bbf} \xbf_2^\dagger \ket{\bbf}... \xbf_{t-1}^\dagger \ket{\bbf} \xbf_t^\dagger \ket{\bbf} }{4^t}  \eta \xbf_t \big( \bigtriangledown f(\xbf_t) \big)^\dagger
\end{align}
Similarly, it takes the same amount to produce the block encoding of
\begin{align}
     \frac{ \big(\xbf_0^\dagger \ket{\bbf} \xbf_1^\dagger \ket{\bbf} \xbf_2^\dagger \ket{\bbf}... \xbf_{t-1}^\dagger \ket{\bbf} \xbf_t^\dagger \ket{\bbf} \big)^\dagger}{4^t}  \eta  \bigtriangledown f(\xbf_t)\xbf_t^\dagger 
\end{align}
and it takes $\mathcal{O}(4s^2)$ block encodings of $\textbf{X}_t$ and $\mathcal{O}(s^2 11)$ block encodings of $A/s,A^\dagger/s$ to build the block encoding of: 
\begin{align}
    & \frac{ \xbf_0^\dagger \ket{\bbf} \xbf_1^\dagger \ket{\bbf} \xbf_2^\dagger \ket{\bbf}... \xbf_{t-1}^\dagger \ket{\bbf} \xbf_t^\dagger \ket{\bbf} }{4^t} \eta^2 \Big(  \big( \Ibb + A^\dagger A \big)  \xbf_t \xbf_t^\dagger \big( \Ibb + A^\dagger A \big)  -  \big( \Ibb + A^\dagger A \big)  \xbf_t \bra{\bbf} A - \ket{\bbf} \xbf_t^\dagger \big( \Ibb + A^\dagger A \big)  + A^\dagger \ket{\bbf}\bra{\bbf} A  \Big) \\
     &= \frac{ \xbf_0^\dagger \ket{\bbf} \xbf_1^\dagger \ket{\bbf} \xbf_2^\dagger \ket{\bbf}... \xbf_{t-1}^\dagger \ket{\bbf} \xbf_t^\dagger \ket{\bbf} }{4^t} \eta^2 \bigtriangledown f(\xbf_t) \bigtriangledown f^\dagger (\xbf_t)
\end{align}
where we remind that we have defined the hyperparameter $\eta = \alpha/8$. Thus in total, it takes 9 block encodings of $\textbf{X}_t$ and $17$ block encodings of $A/s,A^\dagger/s$ to construct the block encoding of:
\begin{align}
   & \frac{ \xbf_0^\dagger \ket{\bbf} \xbf_1^\dagger \ket{\bbf} \xbf_2^\dagger \ket{\bbf}... \xbf_{t-1}^\dagger \ket{\bbf} \xbf_t^\dagger \ket{\bbf} }{4^t} \frac{1}{4}\Big( \xbf_t\xbf_t^\dagger - \eta \xbf_t \big( \bigtriangledown f(\xbf_t) \big)^\dagger - \eta  \bigtriangledown f(\xbf_t)\xbf_t^\dagger + \eta^2 \bigtriangledown f(\xbf_t) \bigtriangledown f^\dagger (\xbf_t)   \Big) \\
 &=  \frac{ \xbf_0^\dagger \ket{\bbf} \xbf_1^\dagger \ket{\bbf} \xbf_2^\dagger \ket{\bbf}... \xbf_{t-1}^\dagger \ket{\bbf} \xbf_t^\dagger \ket{\bbf} }{4^{t+1}} \Big(  \xbf_t-\eta \bigtriangledown f(\xbf_t) \Big) \Big(  \xbf_t-\eta \bigtriangledown f(\xbf_t) \Big)^\dagger \\
 &=  \frac{ \xbf_0^\dagger \ket{\bbf} \xbf_1^\dagger \ket{\bbf} \xbf_2^\dagger \ket{\bbf}... \xbf_{t-1}^\dagger \ket{\bbf} \xbf_t^\dagger \ket{\bbf} }{4^{t+1}}  \xbf_{t+1} \xbf_{t+1}^\dagger
\end{align}
Let $\mathscr{C}(A)$ denotes the circuit complexity of producing the block encoding of $A/s$ (also of $A^\dagger/s$). So the circuit complexity for producing the block encoding of the above operator is 
\begin{align}
   \mathscr{C}(\xbf_{t+1}) &=  \mathcal{O}(s^2) \Big( 9 \mathscr{C}(\xbf_t)  + 17 \mathscr{C} (A)\Big)  \\
   &= \mathcal{O}\Big( 9 \mathscr{C}(\xbf_t)s^2 +  17 \mathscr{C} (A)s^2 \Big)
\end{align}
By replacing $\mathscr{C}(\xbf_t) = \mathcal{O}\Big( 9\mathscr{C}(\xbf_{t-1})s^2  + 17 \mathscr{C}(A)s^2 \Big)$ , we have:
\begin{align}
    \mathscr{C}(\xbf_{t+1}) &=  \mathcal{O}\Big( (9s^2)^2 \mathscr{C}(\xbf_{t-1})   + 9s^2 \cdot17 \mathscr{C}(A)s^2 +  17 \mathscr{C}(A)s^2 \Big)  \\
    &=\mathcal{O}\Big(  (9s^2)^3 \mathscr{C}(\xbf_{t-2})  + (9s^2)^2 \cdot 17s^2 \mathscr{C}(A) +  9s^2\cdot 17 \mathscr{C}(A)s^2 + 17 \mathscr{C}(A) s^2 \Big) \\
    & \vdots \\
    &= \mathcal{O}\Big( (9s^2)^t \mathscr{C}(\xbf_{0})+ \big((9s^2)^{t-1} + (9s^2)^{t-2} + \cdots + (9s^2)^1 + (9s^2)^0 \big) \cdot 17 s^2 \mathscr{C}(A) \Big) \\
    &=  \mathcal{O}\Big( (9s^2)^t \mathcal{O}(1) +  \frac{ (9s^2)^t - 1}{9-1} \cdot 17 s^2\mathscr{C}(A) \Big)
\end{align}
Reminding that $\mathscr{C}(A) = \mathcal{O}\big(\log n+  \log^{2.5} \frac{s}{\epsilon}\big)$ from Lemma \ref{lemma: As}, so for $T$ iteration steps, we have the total complexity is:
\begin{align}
  \mathscr{C}(\xbf_{T})  = \mathcal{O}\Big( (9s^2)^{T+1} \big( \log n + \log^{2.5} \frac{s}{\epsilon}\big)  \Big)
\end{align}
Previously, we have worked out that $\epsilon =  \mathcal{O}\Big( \delta \big(\frac{4}{9}\big)^{T-1}  \Big)$ in order to achieve a final error of $\delta$. In other words, we obtain the $\delta$-approximated block encoding of 
$$ \frac{ \xbf_0^\dagger \ket{\bbf} \xbf_1^\dagger \ket{\bbf} \xbf_2^\dagger \ket{\bbf}... \xbf_{T-1}^\dagger \ket{\bbf} }{4^T} \xbf_T \xbf_T^\dagger $$
So, the total complexity for achieving the above unitary block encoding is $\mathcal{O}\Big( (9s^2)^T \big( \log (n) + (T-1) \log^{2.5} \frac{s}{\delta}  \big) \Big)$. 

Recall that we want to obtain $\ket{\xbf_T}$, which is the quantum state corresponding to the solution. We achieve such a goal by apply $U_T$ as follows:
\begin{align}
        U_T \ket{\bf 0}\ket{\bbf} =  \ket{\bf 0} \Big( \frac{ \xbf_0^\dagger \ket{\bbf} \xbf_1^\dagger \ket{\bbf} \xbf_2^\dagger \ket{\bbf}... \xbf_{T-1}^\dagger \ket{\bbf} }{4^T}  \Big) \xbf_T \xbf_T^\dagger \ket{\bbf} + \ket{\rm Garbage}
\end{align}
Measure the first register and post-select on $\ket{\bf 0}$, then we obtain $\ket{\xbf_T}$. The success probability is:
\begin{align}
    p = \Big| \frac{ \xbf_0^\dagger \ket{\bbf} \xbf_1^\dagger \ket{\bbf} \xbf_2^\dagger \ket{\bbf}... \xbf_{T-1}^\dagger \ket{\bbf}  \xbf_T^\dagger \ket{\bbf}} {4^T}   \Big|^2 |\xbf_T|^2
\end{align}
which will be proved in the Appendix \ref{sec: lowerbound} that it is lower bounded by:
\begin{align}
    p \geq \frac{1}{4^{2T}}  \big|\xbf_0^\dagger \ket{\bbf} \big|^2  \  \big| \xbf_0^\dagger \ket{\bbf} - 3\frac{\alpha T}{8} \big|^{2T} \big| ||\xbf_0|| - 3\frac{T\alpha }{8} \big|^2
\end{align}
In the appendix \ref{sec: lowerbound}, we shall show that $1-\frac{3T\alpha}{8} $ needs to be greater than $0$, and $\xbf_0$ needs to be chosen so that $||\xbf_0|| \leq 1-\frac{3T\alpha}{8 }$. The first condition implies that:
\begin{align}
    \alpha < \frac{8}{3T}
\end{align}
For the second condition we choose $\ket{\xbf_0} = (1-\frac{3T\alpha}{8}) \ket{\bbf}$ for simplicity. In particular, if we choose $\alpha$ such that:
\begin{align}
    \alpha < \frac{4}{6T} < \frac{4}{3T}  \longrightarrow  | 1- 3 \frac{\alpha T}{8} | >  | 1- 6 \frac{\alpha T}{8} | \geq \frac{1}{2}
\end{align}
Then we have:
\begin{align}
    p &\geq   \frac{1}{4^{2T}}  \Big| (1-\frac{3T\alpha}{8})  \Big|^2   \Big| (1-\frac{3T\alpha}{8}) - \frac{3\alpha T}{8} \Big|^{2T}  \Big| (1-\frac{3T\alpha}{8}) - \frac{3\alpha T}{ 8}  \Big|^2  \\ 
   \longrightarrow  p &\geq \frac{1}{4^{2T}}  \frac{1}{4} \big|\frac{1}{2}\big|^{2T} \big|\frac{1}{2} \big|^2
\end{align}
The total complexity for obtaining $\ket{\xbf_T}$ to accuracy $\delta$ is $\mathcal{O}\Big( \frac{(9s^2)^T}{p} \big( \log (n) + (T-1) \log^{2.5} \frac{1}{\delta} \big)  \Big) =\mathcal{O}\Big( 24^{2T} s^{2T}  \big( \log (n) + (T-1) \log^{2.5} \frac{1}{\delta} \big) \Big)$. 

Remind from earlier that as the function $f(\xbf)$ is strongly convex, in order for $\xbf_T$ to be $\delta$-close to the true point of extrema, then according to \cite{nesterov1983method,nesterov2013introductory,boyd2004convex} the choice $T = \mathcal{O}\big( \frac{1}{\eta}\log \frac{1}{\delta}\big) = \mathcal{O}\Big( \log \frac{1}{\delta} \Big)$ is sufficient (where we've used $\eta \sim \frac{1}{\alpha}$ and that $\frac{1}{\eta} = \mathcal{O}(1)$). Thus, we arrive at the following main result:
\begin{theorem}
\label{thm: newlinearsolver}
    Define the linear systems as $A\xbf =\ket{\bbf}$ where $A$ is a real, $s$-sparse Hermitian matrix of size $n\times n$. Assume oracle access to entries of $A$ and a unitary $U_b$ that prepares $\ket{\bbf}$ (also assumed to be real). Let $\ket{\xbf} \sim A^{-1} \ket{\bbf}$ be a quantum state corresponding to solution of $A\xbf = \ket{\bbf}$. Then there exists a quantum algorithm returning a state $\ket{\Tilde{\xbf}}$ such that $ ||  \ket{\Tilde{\xbf} - \ket{\xbf}} || \leq 2\delta$ with complexity 
    $$ \mathcal{O}\Big( s^2 \frac{1}{\delta}  \big( \log(n) + s^2  
 ) \log^{3.5} \frac{s}{\delta} \big) \Big)$$
\end{theorem}
Comparing to \cite{harrow2009quantum} and \cite{childs2017quantum}, which achieves the complexity $\mathcal{O}\Big( s \kappa \frac{\log n}{\delta} \Big)$ and $\mathcal{O}\Big( s \kappa^2  (\log(n) + \rm poly \log \frac{\kappa}{\delta} )\Big)$, our method is free from $\kappa$ -- conditional number of matrix $A$. This is somewhat beyond our expectation. Hence, our method is very suitable for dealing with linear systems with a large conditional number. We note that the relevant work \cite{clader2013preconditioned} was also developed for dealing with a system that has a large conditional number. Compared to that, our method achieves a power-of-five improvement with respect to the sparsity $s$, which is significant. We attribute the major enhancement of our algorithm to the choice of $f(\xbf)$, which is a strongly convex function, enabling a convergence guarantee with a very fast rate. Thus, despite the complexity depends exponentially on $T$ -- the total iteration steps, the value of $T$ can be very small, which results in a efficient final scaling.

\section{Conclusion}
\label{sec: conclusion}
We have successfully developed a new quantum algorithm for solving linear systems of equations. Rather than directly inverting the given matrix, we reformulate the problem as an optimization task, providing an alternative pathway to the solution—specifically, via gradient descent. A key technical distinction of our approach is the mapping of a state vector $\xbf$ to a density operator-like representation $\xbf \xbf^\dagger$. With an appropriately chosen cost function, we derive an analytical expression for the gradient, enabling a straightforward implementation of the gradient descent algorithm to iteratively update the density-operator representation.
We conduct a careful analysis of error accumulation across iterations and the overall time complexity of our framework, revealing that our approach achieves $\kappa$-free scaling, which is an unexpected result. As a trade-off, our method exhibits nearly linear scaling with respect to error tolerance, which appears difficult to improve due to the exponential dependence on T, the total number of iterations. The primary motivation for adopting a density-operator-like representation instead of the conventional vector state representation is to harness a broader range of tools from the quantum singular value transformation (QSVT) framework. Interestingly, we do not rely on any advanced techniques from QSVT but instead utilize only its fundamental definitions and basic block encoding operations. This work thus serves as another compelling example of how QSVT can enhance quantum algorithms, highlighting its vast potential for advancing quantum computing.

\section*{Acknowledgement}
We acknowledge support from the Center for Distributed Quantum Processing at Stony Brook University.

\appendix
\section{Strong convexity of $f(\xbf)$}
\label{sec: strongconvexity}
In this section we elaborate further the strong convex property of:
\begin{align}
    f(\xbf) = \frac{1}{2}||\xbf||^2 +  \frac{1}{2} ||A\xbf||^2  - \bra{\bbf} A\xbf + \frac{1}{2}
\end{align}
We first recall the following definition. A function $f(\xbf): \Rbb^n \longrightarrow \Rbb$ is called $L$-smooth and $\mu$-strongly convex if:\\

$\bullet$ For all $\xbf, \textbf{y} \in \Rbb^n $:
$$f(\xbf) \geq f(\textbf{y}) + \bigtriangledown f(\textbf{y})^T (\xbf - \textbf{y}) + \frac{\mu}{2} || \xbf - \textbf{y} ||^2 $$ 

$\bullet$ For all $\xbf, \textbf{y} \in \Rbb^n $, the gradient of $f(\xbf)$ is Lipschitz with constant $L$: 
$$ || \bigtriangledown f(\xbf) - \bigtriangledown f(\textbf{y}) || \leq L ||\xbf - \textbf{y}||  $$

It turns out that the above conditions are equivalent to Hessian of $f(\xbf)$ is bounded from below and above, respectively \cite{nesterov1983method, nesterov2013introductory, boyd2004convex}. The gradient of function $f(\xbf)$ is $\bigtriangledown f(\xbf) = \xbf +A^\dagger (A\xbf - b)$. The Hessian is $\bigtriangledown^2 f(\xbf) = \Ibb +A^\dagger A$. It is apparent that:
\begin{align}
    || \bigtriangledown^2 f(\xbf) || &= || \Ibb + A^\dagger A || \\
                        &\leq  1 + \lambda_{\max} (A^\dagger A)  \\
                        &\leq  1 + \sigma^2_{\max} (A) 
\end{align}
where $\lambda_{\max}, \sigma_{\max}$ refers to maximum eigenvalue of $A^\dagger A$ and singular value of $A$, respectively. Thus, the Hessian of $f(\xbf)$ is bounded from above, so it is $L$-smooth with $L = 1 + \sigma_{\max}^2 (A)$ . To show that the Hessian is $\mu$-strongly convex, we need to show that it is bounded from below. It is straightforward to see that:
\begin{align}
     || \bigtriangledown^2 f(\xbf) || &\geq 1 + \lambda_{\min} (A^\dagger A) \\
     &\geq 1 + \sigma_{\min}^2 (A)
\end{align}
Thus, $f(\xbf)$ is $\mu$-strongly convex with $\mu = 1 + \sigma_{\min}^2 (A)$.

\section{Evaluation of function $f(\xbf)$}
\label{sec: evaluationfunctionf}
In this appendix we complete what we left in Section \ref{sec: mainframework}, where we mentioned that the ability to prepare $C\xbf$ (where $C$ is some possibly normalization factor), $\ket{\bbf}$ and oracle access to $A$ can be used to estimate the function $f(\xbf)$, which is defined as:
\begin{align}
    f(\xbf) = \frac{1}{2}||\xbf||^2 +  \frac{1}{2} ||A\xbf||^2  - \bra{\bbf} A\xbf + \frac{1}{2}
\end{align}
Given the ability to prepare $C\xbf$, it is simple to use Hadamard/SWAP test to estimate $|C|^2 ||\xbf||^2$, which yields the value of $|\xbf|^2$. Now we use Lemma \ref{lemma: As} to block-encode $A/s$, denoted as $U_A$. According to Definition \ref{def: blockencode} and Eqn.~\ref{eqn: action}, we have:
\begin{align}
    U_A \ket{\bf 0} C\xbf = \ket{\bf 0} \frac{A}{s} C\xbf + \ket{\rm Garbage}
\end{align}
Using amplitude estimation \cite{brassard2002quantum, manzano2023real, rall2021faster,rall2023amplitude} to estimate the amplitude $||  \frac{A}{s} C\xbf|| = \frac{|C|}{s} ||A\xbf ||$, which yields the value of $||A\xbf||^2$. Now we prepare $\ket{\bbf}$, and observe that the overlap:
\begin{align}
    \bra{\bf 0}\bra{\bbf} U_A \ket{\bf 0} C\xbf  &=\bra{\bf 0}\bra{\bbf} \Big ( \ket{\bf 0} \frac{A}{s} C\xbf + \ket{\rm Garbage} \Big) \\
    &= \bra{\bbf}\frac{C}{s} A\xbf = \frac{C}{s} \bra{\bbf} A\xbf
\end{align}
which can be estimated by Hadamard test. Thus the value of $\bra{\bbf} A\xbf $ can be revealed as we know the value of $C$ and $s$. 

\section{Evaluation of $\xbf^\dagger \ket{\bbf} $}
\label{sec: evaluationinnerpdouct}
Suppose we have a unitary block encoding $U_x$ of some $C \xbf \xbf^\dagger$. Recall that we had the unitary $U_b$ that generates $\ket{\bbf}$. According to Definition \ref{def: blockencode} and Eqn.~\ref{eqn: action}, we have:
\begin{align}
    U_x \ket{\bf 0}\ket{\bbf} = \ket{\bf 0} \big(  C\xbf \xbf^\dagger \ket{\bbf} \big) + \ket{\rm Garbage}
\end{align}
Define $\ket{\Phi_1} \equiv \ket{\bf 0} \big(  C\xbf \xbf^\dagger \ket{\bbf} \big) + \ket{\rm Garbage} $, and $\ket{\Phi_2} \equiv \ket{\bf 0}\ket{\bbf}$. The overlap between two states is:
\begin{align}
    \braket{\Phi_1, \Phi_2} &= \bra{\bf 0}\bra{\bbf} \Big( \ket{\bf 0} \big(  C\xbf \xbf^\dagger \ket{\bbf} \big) + \ket{\rm Garbage}  \Big) \\
    &= C \bra{\bbf}\xbf \xbf^\dagger \ket{\bbf} \\
    &= C |\xbf^\dagger \ket{\bbf}|^2
\end{align}
which can be estimated via Hadamard/SWAP test, as we know the unitaries that generate both states. The cost of estimating the above quantity to an accuracy $\delta$ is $\mathcal{O}(\frac{1}{\delta})$.  We remark that $C\xbf \xbf^\dagger \ket{\bbf} = C \xbf^\dagger \ket{\bbf} \  |\xbf| \ket{\xbf} $, thus the sign of $\xbf^\dagger \ket{\bbf}$ can be inferred by performing amplitude estimation \cite{brassard2002quantum, manzano2023real,suzuki2020amplitude,rall2023amplitude,rall2021faster} on the state $\ket{\Phi_1}$. 

In our algorithm, we begin with a block encoding of $\xbf_0 \xbf_0^\dagger$. Then using the above procedure allows us to estimate $| \xbf_0^\dagger \ket{\bbf} |^2$, which can infer the value of $|\xbf_0^\dagger \ket{\bbf}|$. Combining with the sign of $\xbf_0^\dagger \ket{\bbf}$ derived above, the value of $\xbf_0^\dagger \ket{\bbf} $ is known. After the first iteration, we have the block encoding of $\frac{\xbf_0^\dagger \ket{\bbf}}{4} \xbf_1 \xbf_1^\dagger $. Use the same procedure as above, we can estimate the sign of $\frac{\xbf_0^\dagger \ket{\bbf}}{4} \xbf_1^\dagger \ket{\bbf}$ and the magnitude of $\frac{\xbf_0^\dagger \ket{\bbf}}{4} |\xbf_1^\dagger \ket{\bbf} |^2 $. Given that the sign and value of $\xbf_0^\dagger \ket{\bbf}$ is known, the value of $ |\xbf_1^\dagger \ket{\bbf}|^2$ can be estimated and infer the value and sign of $\xbf_1^\dagger \ket{\bbf}$. After the second iteration, we have the block encoding of $\frac{\xbf_0^\dagger \ket{\bbf} \xbf_1^\dagger \ket{\bbf} }{4^2} \xbf_2 \xbf_2^\dagger  $. Using the same procedure, we can find the sign of $\frac{\xbf_0^\dagger \ket{\bbf} \xbf_1^\dagger \ket{\bbf} }{4^2} \xbf_2^\dagger \ket{\bbf} $ and magnitude of $ \frac{\xbf_0^\dagger \ket{\bbf} \xbf_1^\dagger \ket{\bbf} }{4^2} |\xbf_2^\dagger \ket{\bbf}|^2 $. The values and signs of $\xbf_0^\dagger \ket{\bbf}$ and $\xbf_0^\dagger \ket{\bbf}$ are known, thus the value and sign of $\xbf_2^\dagger \ket{\bbf}$ can be inferred. After $t$ iterations, we have the block encoding of:
$$\frac{ \xbf_0^\dagger \ket{\bbf} \xbf_1^\dagger \ket{\bbf} \xbf_2^\dagger \ket{\bbf}... \xbf_{t-1}^\dagger \ket{\bbf}  }{4^{t}} \xbf_t \xbf_t^\dagger $$
Using the same method, we can determine the value and sign of $\xbf_t^\dagger \ket{\bbf}$.

\section{Lower bound for success probability p}
\label{sec: lowerbound}
Remind that the success probability is:
\begin{align}
    p = \Big| \frac{ \xbf_0^\dagger \ket{\bbf} \xbf_1^\dagger \ket{\bbf} \xbf_2^\dagger \ket{\bbf}... \xbf_{T-1}^\dagger \ket{\bbf}  \xbf_T^\dagger \ket{\bbf}} {4^T}   \Big|^2 ||\xbf_T||^2
\end{align}
We also have the gradient of $f(\xbf)$:
\begin{align}
    \bigtriangledown f(\xbf) = \big( \Ibb + A^\dagger A \big) \xbf - A^\dagger \ket{\textbf{b}}
\end{align}
Thanks to triangle inequality, we have:
\begin{align}
    ||\bigtriangledown f(\xbf)|| \leq || \big( \Ibb + A^\dagger A \big) \xbf || + || A^\dagger \ket{\textbf{b}}||
\end{align}
For any $\xbf$ such that $||\xbf|| \leq 1$:
\begin{align}
    || \big( \Ibb + A^\dagger A \big) \xbf ||  \leq ||\big( \Ibb + A^\dagger A \big)   || \leq 1 + 1 = 2
\end{align}
where we have used that $||A|| \leq 1$, which also implies that $||A^\dagger \ket{\bbf}|| \leq 1$. So, the norm of gradient $||\bigtriangledown f(\xbf) || \leq 3$. At $t$-th iteration, we have:
\begin{align}
    \xbf_t &= \xbf_{t-1}  - \eta \bigtriangledown f(\xbf_{t-1}) \\
    &= \xbf_{t-2} - \eta \bigtriangledown f(\xbf_{t-2}) - \eta \bigtriangledown f(\xbf_{t-1}) \\
    & \vdots \\
    &= \xbf_0 - \eta \bigtriangledown f(\xbf_0) - ... -  \eta \bigtriangledown f(\xbf_{t-2}) - \eta \bigtriangledown f(\xbf_{t-1}) 
\end{align}
taking inner product, we have:
\begin{align}
    \bra{\bbf} \xbf_t = \bra{\bbf}\xbf_0 - \eta \bra{\bbf}   \bigtriangledown  f(\xbf_0) - ... -  \eta \bra{\bbf}   \bigtriangledown f(\xbf_{t-2}) - \eta  \bra{\bbf}  \bigtriangledown \bra{\bbf}f(\xbf_{t-1}) \\
    \longrightarrow |\bra{\bbf} \xbf_t |^2 = |\bra{\bbf}\xbf_0 - \eta  \bra{\bbf}  \bigtriangledown  f(\xbf_0) - ... -  \eta \bra{\bbf}  \bigtriangledown  f(\xbf_{t -2}) - \eta  \bra{\bbf} \bigtriangledown f(\xbf_{t -1}) |^2
\end{align}
It is apparent that for any $j=0,1,2,...,T-1$:
\begin{align}
    \bra{\bbf} \bigtriangledown f(\xbf_j) \leq |\bigtriangledown f(\xbf_j)|  \leq 3
\end{align}
Thus:
\begin{align}
    |\bra{\bbf} \xbf_t |^2 \geq | \bra{\bbf} \xbf_0  - \eta 3t  |^2 \geq | \bra{\bbf} \xbf_0 - \eta 3 T|^2
\end{align}
where in the last inequality we have used $t \leq T$. Additionally, we also have:
\begin{align}
   \xbf_t =\xbf_0 - \eta \bigtriangledown f(\xbf_0) - ... -  \eta \bigtriangledown f(\xbf_{t-2}) - \eta \bigtriangledown f(\xbf_{t-1}) \\
   \longrightarrow ||\xbf_t||^2 \geq \Big| ||\xbf_0|| - \eta ||\bigtriangledown f(\xbf_0)|| - \eta ||\bigtriangledown f(\xbf_1)||- ... - \eta ||\bigtriangledown f(\xbf_{t-1})||  \Big|^2 \\
   \longrightarrow ||\xbf_t||^2 \geq \Big| ||\xbf_0|| - 3\eta T  \Big|^2
\end{align}
Combining everything, we have:
\begin{align}
    p &= \Big| \frac{ \xbf_0^\dagger \ket{\bbf} \xbf_1^\dagger \ket{\bbf} \xbf_2^\dagger \ket{\bbf}... \xbf_{T-1}^\dagger \ket{\bbf}  \xbf_T^\dagger \ket{\bbf}} {4^T}   \Big|^2 ||\xbf_T||^2 \\
    &\geq \frac{1}{4^{2T}} \big|\xbf_0^\dagger \ket{\bbf} \xbf_1^\dagger \ket{\bbf} \xbf_2^\dagger \ket{\bbf}... \xbf_{T-1}^\dagger \ket{\bbf}  \xbf_T^\dagger \ket{\bbf} \big|^2 ||\xbf_T||^2 \\
    & \geq  \frac{1}{4^{2T}}  |\xbf_0^\dagger \ket{\bbf}|^2 | \xbf_0^\dagger \ket{\bbf} - 3\eta T|^{2T} \big| ||\xbf_0|| - 3\eta T  \big|^2
\end{align}
Remind that we have chosen $\eta = \alpha /8$, so we have:
\begin{align}
    p \geq \frac{1}{4^{2T}}  \big|\xbf_0^\dagger \ket{\bbf} \big|^2  \  \big| \xbf_0^\dagger \ket{\bbf} - 3\frac{\alpha T}{ 8} \big|^{2T} \big| ||\xbf_0|| - 3\frac{\alpha T}{8} \big|^2
\end{align}
We remark that an important property that we have used is that $|| \bigtriangledown f(\xbf) || \leq 3$ for $\xbf$ satisfying $||\xbf|| \leq 1$. We need to guarantee that for the whole iterations, $||\xbf_0||, ||\xbf_1||, ..., ||\xbf_T|| \leq 1$. We recall that:
\begin{align}
    \xbf_T= \xbf_{T-1} - \eta \bigtriangledown f(\xbf_{T-1})
\end{align}
Using triangle inequality, we have:
\begin{align}
    ||\xbf_T|| \leq ||\xbf_{T-1}|| + \eta || \bigtriangledown f(\xbf_{T-1})||
\end{align}
By induction, we have that $||\xbf_T|| \leq ||\xbf_0|| + \eta ||\bigtriangledown f(\xbf_0)|| + ... +  \eta ||\bigtriangledown f(\xbf_{T-1}) ||$. Since we want $||\xbf_t|| \leq 1$ for all $t$, then it suffices to have:
\begin{align}
    ||\xbf_0|| + \eta ||\bigtriangledown f(\xbf_0)|| + ... +  \eta ||\bigtriangledown f(\xbf_{T-1}) || \leq 1 \\
    \longrightarrow ||\xbf_0|| + 3 \eta T \leq 1 \\
    \longrightarrow ||\xbf_0|| + 3 T \frac{\alpha }{8 }  \leq 1 
\end{align}
Thus we need to choose $\xbf_0$ such that $||\xbf_0|| \leq 1- 3 T \frac{\alpha }{8}  $. We first need to guarantee that $ 0< 1- 3 T \frac{\alpha }{8}    $, which implies:
\begin{align}
    \alpha \leq \frac{8}{3T}
\end{align}
In the main text, we chose $\alpha < \frac{4}{3T}$, which naturally satisfies the above condition. Hence, for a total $T$ iterations, it is guaranteed that $||\xbf_t|| \leq 1$ for $t=0,1,2,...,T$.

\bibliography{ref.bib}{}
\bibliographystyle{unsrt}

\clearpage
\newpage
\onecolumngrid

\end{document}